\documentclass[aps,epsfig,preprint,groupedaddress]{revtex4-1}
\usepackage{amssymb}
\usepackage{graphicx}
\usepackage{color}
\usepackage{hyperref}
\usepackage{fullpage}
\usepackage{comment}
\makeatletter
\gdef\@ptsize{0}
\let\@currsize\normalsize
\makeatother
\begin{document}
\title {Effect of Heterogeneous Transmission Rate on Epidemic Spreading Over Scale Free Networks}
\author{Vikram Sagar, Yi Zhao}
\affiliation{Shenzhen Graduate School, Harbin Institute of Technology, Shenzhen 518055 ,China}
\email{zhao.yi@hitsz.edu.cn}
\email{sagar@hitsz.edu.cn}
\begin{abstract}
 In the present work the spread of epidemic is studied over complex networks which are characterized by power law degree distribution of links and heterogeneous rate of disease transmission. The random allocation of epidemic transmission rates to the nodes results in the heterogeneity, which in turn causes the segregation of nodes in terms of various sub populations. The aim of the study is to gain microscopic insight into the effect of interactions among various sub populations in the spreading processes of disease over such networks. The discrete time Markov chain method based upon the susceptible infected susceptible (SIS) model of diseases transmission has been used to describe the spreading of epidemic over the networks. The study is parameterized in terms of variable $\lambda$, defined as the number of contacts a node makes with the fraction of its neighboring nodes. From the simulation results, it is found that the spread of epidemic on such networks is critical in terms of number of minimum contacts made by a node below which there is no outbreak of disease. The degree of infection in these networks is assessed from the size of epidemic defined in terms of fraction of infected nodes of the total number and their corresponding level of infection. The results of the parametric study demonstrates the dependence of the epidemic size upon number of concurrent contacts made by a node ($\lambda$ ) and the average number of links per node. In both these cases, the size of the epidemic is found to increase with the corresponding increase in respective parameters.
\end{abstract}
\maketitle

\section{Introduction}
The topic of epidemic spreading has been of great interest for very long period of time\cite{histro-1} owing to its application in the multitude of research areas\cite{app-1,app-2,app-3,app-4,app-5,app-6}. The intensive research activity in the area of complex networks over the last few decades has turned them into a very conducive substrate to study a plethora of problems pertaining to the spread of epidemics\cite{histro-2,histro-3,histro-4,histro-5}. The general prescription to study these problems in the realms of complex network theory involves representing the networks such as biological, social, technical or financial in terms of graphs and their subsequent coupling with suitable spreading process\cite{app-3,app-4,app-5,math-1,math-2,math-3}. In this description, the entities forming a network such as humans, computers, airports, biological cells, emails, internet etc. are represented in terms of nodes and the corresponding interactions between them as links. The networks can be categorized into different kinds based upon the structural organization of their constituent nodes and the corresponding links connecting them. In the literature a special interest has been in the study of scale free networks\cite{BA-1,BA-2,SF-1} for which the connectivity between the links is described in terms of power law distribution given by $P(K)\sim K^{-\gamma}$, where the parameter $\gamma$ must be larger than zero to ensure a finite average connectivity $<K>$. The power law distribution for the links implies that each of the nodes in the network have statistically significant probability of having very large number links in comparison to the average value. The scale free characteristics have been observed in many of the naturally and artificially occurring networks\cite{SF-1,SF-2,SF-3}.

 The susceptible-infected-susceptible (SIS) is a simple model that has served as a building block to study vast number of problems pertaining to the spread of epidemics on these networks both at macroscopic as well as microscopic scales\cite{app-5,SIS-6,SIS-7}. In this model the nodes of a network are either in susceptible or are in infected state. The infection spreads when a susceptible node comes in contact with infected node which results in the change of state of the latter. This change of state from susceptible to infected is at a rate proportional to the number of contacts times the rate of spread ($\beta$). On the other hand the infected individual recovers to the susceptible state at a recovery rate $\mu$.  The macroscopic description of epidemic spreading upon these networks rely upon the mean field approach in which the homogeneity and isotropy of the system are used to reduce the level of complexity. The extension to the previous approach is a heterogeneous mean field approach in which the nodes are segregated into different classes depending upon their respective degrees and the dynamical properties are assumed to be same for each of these classes. This coarse grained macroscopic description has been successful in the study of critical properties of the network, such as outbreak and prevalence of epidemics\cite{SIS-1,SIS-2,SIS-3,SIS-4,SIS-5}. The understanding of topological effects of network upon the spreading of epidemics as well as the node level probabilistic study of the mechanism are in the preview of the microscopic description\cite{app-4,SIS-6,SIS-7,SIS-8}.

   In these aforementioned studies epidemic spreading has been assumed to have a homogeneous transmission with a constant spreading rate ($\beta$) and the heterogeneity is considered in terms of the  number of links connected to the node. However, the real scenarios representing the social or computer networks can be much more complicated in which multiple contagions can be simultaneously  present and their effects are not mutually exclusive. For such a case, single source heterogeneity is not suffice to describe the spread of a epidemic. The complex coupling arising from the multiple heterogeneities have to be taken into account to describe the possible synergetic impact of multiple contagions on the disease transmission\cite{intro-8}. Alternatively, it can be of special interest in understanding the spread of same disease in social networks that comprises different sub-populations which can be categorized on the basis of age, social status, sex or different levels of immunity. The spread of disease on such networks is not only dependent upon the degree of the node but also upon the individual rate of transmission.

The present work takes into account the effects of heterogeneous rates of infection transmission upon the spreading dynamics of the disease over scale free networks \cite{BA-1,BA-2}. The Monte Carlo Markov chain method based upon SIS model is used to describe the spread of disease over these networks. It provides the first principle node level description of the infection mechanism parameterized in terms of the number of contacts a node makes with the fraction of its neighboring nodes at each time step. The parameterization in terms of number of concurrent contacts made by a node to its neighbors has been used to model the different spreading processes\cite{SIS-6,SIS-7}. The two of the prominent models based upon this criterion are: Contact Process (CP) and Reactive Process (RP). These two process represent the limiting cases as in the former case (CP), a node makes contact with only one of its neighbor and in the latter case (RP), a  node makes simultaneous contact with all its neighbors. In this work, the time evolution of the nodes over a parametric space defined in terms of rate of infection ($\beta$), node degree (number of links $K$) and the state of the system ($\rho(t)$) is examined to understand the interplay between the network topology and criticality of the parameters in the outbreak of disease. The size of an epidemic on infected network is measured  in terms of the fraction of infected individuals and the corresponding level of their infection. The epidemic size is determined from the final(or equilibrium) state of the network at which the activity of all the nodes(or infection level) becomes stationary. The explicit dependence of the epidemic size upon each of these previously stated factors is studied by projecting them in their respective ($\beta-\rho$) and ($K-\rho$) parametric spaces. This detailed description gives deeper insight into the interdependence between network topology and spreading processes over the final size of epidemic.

 Further the reactive processes (RP) in which an infection spreads by the simultaneous contact from infected node to all its neighbors are explicitly studied on a network with size N. For this process, the dependence of the final epidemic size on the network topology together with the time of spread of infection can be of great interest for design of vaccination strategies. Thus a comparative study has been carried out to understand the dependence of epidemic size and the spreading time of infection on the network topology parameterized in terms of minimum degree of the nodes (i.e different $m(=1,2,3)$ values of BA model).

 The organization of the paper is as follows:  In section (II), theoretical frame work for studying the spread of disease over the finite sized scale free network with heterogeneous transmission rate is presented.  In section (III), the results of the Monte-Carlo simulations are presented highlighting the interdependence of various parameters together with the network topology upon outbreak, prevalence and size of epidemic.  Section (IV), contains the summary and discussion of the results.

\section{Theoretical Framework}
A brief introduction to the susceptible-infected-susceptible (SIS) model is presented in this section that has been used to describe the spread of disease over the network. It belongs to a class of compartmental models, in which the individuals are partitioned in different compartments or states depending upon the stage of infection. In this model, an individual can be in either of the two states namely susceptible (denoted by {\bf S}, those which are prone to infection) or infected (denoted by {\bf I}, those which are carrying the infection and are thus contagious). The model is paradigmatic as the microscopic details of the mechanism causing the infection to an individual are ignored and the state is assessed only macroscopically. The state of an individual in this model undergoes stochastically looped over transitions between the following three states namely {\bf S}(susceptible) $\rightarrow$ {\bf I}(infected) $\rightarrow$ {\bf S}(susceptible) till it attains a stationary value. The network description of the population involves representing the individuals in terms of the nodes and the mutual connections between them as the edges. The adjacency matrix $A=[a_{ij}]_{N\times N}$ of the  network is a binary valued matrix in which the presence or absence of connection between a pair of nodes $(i,j)$ is by given by $a_{ij}=1$ and $a_{ij}=0$ respectively. For the present work, the edges(or connections) between the N nodes(or individuals) are unweighted and undirected with a power law degree distribution $P(k)\sim k^{-\gamma}$(where $\gamma >0$) based upon Barabasi Alberta (BA) Model\cite{BA-1,BA-2}. In this work, the nodes of the network are characterized in terms of two parameters namely the node degree ($K$) and rate of transmission ($\beta$). The degree of the node is described by a power law distribution ($K$) and its corresponding transmission rate ($\beta_{i}$) is randomly assigned a discrete value from a specified range $\{.1,1\}$. The rates of transmission are discretized  (in units of $0.1$) and they have been normalized to the rate recovery $\mu$ which is set to a numerical value one. The random allocation of the transmission rate makes the spreading of disease heterogeneous, as the infection probability between a pair of nodes ($i,j$) is not equal. The schematic layout of the model is shown in Figure-1, to describe mechanism of heterogeneous spread of infection by a node to its neighbors.

 The mechanism for the spread of disease over the network based upon SIS model can be physically understood in the following way. An infected node in the network transmits the infection with a probability $\beta$ to its immediate neighborhood. The successful transmission to an initially susceptible node results in its change of state to become infected, and a successful transmission to an initially infected node has no effect. An infected node can get cured and change its state to susceptible with a probability $\mu$. The spread of the disease follows a Markovian assumption as the successive states of each the nodes is only affected by their present state. The successive synchronous transitions of the system to a stationary system results in Markov chains. Based upon this Markovian assumption a unified model has been proposed in which various spreading process are parameterized in terms of number of concurrent contacts a node makes with its neighbors and attempts to transmit the disease\cite{SIS-6,SIS-7}. Thus a node can concurrently receive infection from many other nodes in the network and change its state according to above described mechanism. In the present work, this model has been extended to take into the heterogeneous rate of disease transmission into account. The following is the discrete-time stochastic mathematical model of the above described process.
\begin{eqnarray}
  p_{i}(t + 1) &=& (1-q_{i}(t)) (1-p_{i}(t)) + (1-\mu)p_{i}(t)+\mu (1-q_{i}(t))p_{i}(t), \label{eq-1}
\end{eqnarray}
The eq-(\ref{eq-1}) can be re-arranged and expressed in the simplified form,
\begin{eqnarray}
  p_{i} (t + 1) &=& (1-q_{i}(t)) + (1-\mu)p_{i}(t)q_{i}(t),\label{eq-2}
\end{eqnarray}
where the probability $q_{i}(t)$ of node $i$ not being infected by any neighbor is given by,
\begin{eqnarray}
  q_{i}(t) &=&\prod_{j=1}^{N}[1-\beta_{i}r_{ji} p_{j}(t)]. \label{eq-3}
\end{eqnarray}
The random allocation of the transmission rate described by $\beta_{i}$ to the $i^{th}$ node of network introduces the two parameter heterogeneity ($\beta_{i},K_{i}$) in the system.  The role of the individual terms constituting the  Eq.(\ref{eq-1}) can be understood from the model description outlined in the following references\cite{SIS-6,SIS-7}. The probability that a susceptible node $(1-p_{i}(t))$ is infected $(1-q_{i}(t)$ by at least a neighbor is given by the first term. The second term gives the probability that node infected at time $t$ does not recover $(1-\mu)p_{i}(t)$ , and finally the last term takes into account the probability that an infected node recovers $(\mu p_{i}(t))$ but is re-infected by at least a neighbor $(1-q_{i}(t)$. Within this formulation, it is assumed that the most general situation in which recovery and infection occur on the same time scales, allowing then re-infection of individuals during a discrete time window (for instance, one MC step).
The values of the contact probabilities $r_{ji}$ for unweighed networks can be expressed as,\\
\begin{equation}\label{eq-4}
  r_{ji} = R_{\lambda} (\frac {a_{ij}}{k_{j}})= a_{ji}R_{\lambda}(\frac {1}{k_{j}})= a_{ji}R_{\lambda}(k_{j}^{-1})
\end{equation}
 where
 \begin{eqnarray}\label{eq-5}
R_{\lambda}(x)&=& 1-(1-x)^{\lambda}.
\end{eqnarray}
For the contact process, $\lambda = 1$ and $R_{1}(x) = x$, whereas for the fully reactive process, $\lambda \rightarrow \infty $ and $R_{\infty}(x) = 1$. At the stationary state, Eqs.(\ref{eq-1}) and (\ref{eq-3}) are independent of the discrete time step, and simplify to
\begin{eqnarray}
  p_{i}  &=& (1-q_{i}) + (1-\mu)p_{i}q_{i}\label{eq-6}
\end{eqnarray}
with
\begin{eqnarray}
  q_{i} &=&\prod_{j=1}^{N}[1-\beta_{i}r_{ij} p_{j}]. \label{eq-7}
\end{eqnarray}
The final stationary state of the system can be also be determined solving the Eqs.\ref{eq-6} and \ref{eq-7} by fixed-point iteration until a fixed point is found. Finally, the average fraction of infected nodes in the stationary state is given by
\begin{eqnarray}
  \rho_{av} &=&\frac{1}{N} \sum_{j=1}^{N}\rho_{j}. \label{eq-8}
\end{eqnarray}

 The simultaneous dependence of the infected node upon the transmission rate ($\beta_{i}$) and node degree ($K_{i}$), which are not mutually exclusive hinders the previously outlined mathematical frame work \cite{SIS-6} for determining the criticality of parameter in the  outbreak of a disease.

\section{RESULTS}
In this section, the results of MC simulations are presented so as to gain detailed insight into the mechanism of disease spread over scale free networks with the heterogeneous rate of transmission.  The spread of disease over these networks is based upon the previously described SIS model in which a given node at any time can be in either of the two states namely susceptible ($S$) or infected ($I$).  The BA model has been used to generate the scale free networks with the power law degree distribution of the links with the value of exponent ($\gamma$) in the range of $2.73\sim3.059$. In these simulations, the number of the nodes ({\bf N}) in the network are kept constant ($2\times10^3$). The study of different spreading processes is parameterized in terms of variable $\lambda$, which denotes is the number of contacts a node makes with the fraction of its neighboring nodes. This parameterization scheme provides a comprehensive quantitative understanding and simultaneous comparison of various disease spreading process in the networks such as contact($\lambda=1$), random walker($2\leq \lambda < \infty $) and reactive($\lambda=\infty$). At the onset of the simulation a certain fraction of the nodes ($5\%$) are randomly infected. The coupled set of equations given by Eqs.-(\ref{eq-1}) and (\ref{eq-3}) describe the dynamical state of a network at any given time. These equations are simultaneously solved to describe the spread of the disease on the networks and are evolved till the state of nodes attain a stationary value (i.e $\rho(i)(t+1)-\rho(i)(t)\sim 0$).

In Figure 2, the detailed phase space diagrams describing various spreading processes are shown in the parametric space defined by the node degree($K$), rate of transmission($\beta$) and normalized final state ($\rho_{f}$). The phase space diagram highlights the interaction between the different spreading processes (parameterized in terms of $\lambda$) and the network topology($K,\beta$) on the spread of disease. From the results, it can inferred that the spread of epidemic on such networks displays a critical behavior in terms of the fraction of minimum number of concurrent contacts made by a node with its neighbor. For the results corresponding to $\lambda=1$, it is evident that there is no outbreak of epidemic for a contact process (CP). The results for intermediate cases ($\lambda=3,5$) and reactive process ($\lambda=\infty$) are shown in the subsequent sub plots of the figure. It is further evident from the results, that the final state of respective nodes of the network in the parametric space has complex a form. The complexity arises due to simultaneous dependence of final state upon the rate of transmission of the respective node and their corresponding degree of node.

The degree to which an infection spreads in the networks corresponding to various spreading processes (for different $\lambda$) can be assessed in terms of size of epidemic. In the present study, the size of epidemic is expressed in terms of the fractional distribution of infected individuals of the total population ($dN/N$) together with their subsequent level of infection($\rho_{f}$). In Figure 3, the results of different parametric studies are shown which gives the relative quantitative measure of epidemic sizes for each of the spreading processes that are parameterized in terms of variable $\lambda$. From the simulated results, it has been found that for a class of spreading processes described by $\lambda \geq 2$, there is an outbreak and then the prevalence of disease over these networks. It can be further inferred from such results that the size of epidemic over a network heavily relies on the choice of spreading processes. The size of epidemic is found to increase with the increase in parameter $\lambda$, which refers to variation from contact processes ($\lambda=1$) towards the reactive processes($\lambda_{10}=\infty$). The increase in the epidemic size can be attributed to the increased participation of the individuals described by Eq-\ref{eq-4} for the various spreading processes. Physically, it signifies increased level of interaction among the individuals in the connected network. It is worthwhile to mention here that for representational purpose the results of reactive process (i.e, $\lambda=\infty$) have been plotted at $\lambda=10$ and doesn't correspond to the latter parameter.

   In Figure 4, the explicit dependence of the epidemic sizes upon each of the two factors i.e, rate of transmission ($\beta_{i}$) and corresponding node degree ($K$) is shown in the respective parametric spaces ($\beta-\rho$ $\&$  $K-\rho$) for intermediatory spreading processes($\lambda=3,5$). These findings address the role of collective dynamics exhibited by the different sub populations during the spread of infection. It also highlights the microscopic criticality for the outbreak of the disease for a specified spreading process. The understanding of this microscopic criticality can be effective in the efficient design of vaccination program for the eradication of the disease by individualized targeting of sub populations. The comparison between the different epidemic sizes in the $\beta-\rho$ parametric space shows the dependence of the epidemic size upon corresponding transmission rate of infection($\beta_{i}$). Similarly, the other evident result is increase  in the epidemic size in $K-\rho$ parametric space with the value of parameter $\lambda$ corresponding to different spreading processes. This fine grained categorization of the nodes in terms of two parameters indicates the difficulty in the analytical approach for the study of system properties.

The another aspect of the study is to understand the spread of infection corresponding to the reactive process which is one of the widely studied mechanisms of infection transmission. The reactive processes represents the upper limit ($\lambda=\infty$) of the above described unified model for different processes. In this scenario, an infected node concurrently contacts all its nearest neighbors and attempts to spread the infection to them. In Figure 5, the resultant epidemic sizes corresponding to a reactive process over the scale free networks with heterogeneous rate of transmission are demonstrated. The resultant epidemic sizes correspond to different values of parameter $m(=1,2,3)$ of the BA model. The value of parameter $m$ is indicative of minimum degree of the node in a network. It is evident from the results that the epidemic size depends upon the minimum degree of the node in a network(i.e network topology) and the increase in the epidemic size is proportional to the value of parameter $m$.  This can be inferred from the shrinking of epidemic spread corresponding to the different values of variable $m$ and the gradual shift of the epidemic size towards the maximum value.

   The explicit dependence of each of these epidemic sizes on parameters $\beta_{i}$ and $K_{i}$ in a reactive process corresponding to different networks is shown in Figure 6. It is evident from the results that the epidemic size has non zero value in each of these spaces which implies the absence of microscopic threshold about these parameters in the outbreak of the epidemic. The comparison of the relative epidemic sizes in the $\beta-\rho$ and $K-\rho$ spaces demonstrates the underlying role of network topology upon the interactional dynamics of the sub populations. In the $\beta$ space, for a specified value of parameter $m$, the relative epidemic size for different sub populations is proportional to the respective rate of transmission ($\beta_{i}$). Further the increase in the value of parameter $m$ results in the relative increase in epidemic size for various sub populations. The similar dependence is also observed in the relative comparison between epidemic size in the $K$ space. The increase in the value of parameter $m$ results in the shift of population towards larger epidemic size.

 The rate at which the epidemic spreads along with its size can be of great relevance for the design of efficient vaccination program. In the present work the velocity at which an epidemic spread for a unit time step can been approximately defined as: $\dot{\rho}(t)\simeq \rho(t+1)-\rho(t)$. In Figure 7, the time evolution of epidemic size and the corresponding velocity are shown for a different values of parameter $m(=1,2,3)$.  It is evident from the results that the time of spread of epidemic depends upon minimum degree of the node that is parameterized in terms of previously defined variable $m$. The increase in number of links in a networks corresponding to the minimum degree of node increases the rate of spread of epidemic. Thus it can be conjectured that the time of spread and size of the epidemic for the reactive process depends upon the minimum node degree ($m$).

\section{Summary}
 The theoretical framework together with the results of the present study are summarized in this section. The framework presented here has wide range of applications in the description of various real life scenarios, such as to understand the role of interaction among various sub-populations in a given sample in the spread of the disease. It can also be used to describe the effect of simultaneous presence of multiple contagions in the social or computer networks having different probability of epidemic transmission. In the present work, the spreading of disease is studied using SIS model over scale free networks that are characterized by heterogeneous rates of infection transmission. The scale free networks are described by BA model and degree distribution of the links is given by $P(k)\sim k^{-\gamma}$ with $\gamma$ in the range $2.73 \sim 3.01$. The spread of disease over these networks is described using a discrete time MC simulations in which the state of all the nodes in the network is synchronously updated at each time step. In these simulations the state of all the nodes in the network is evolved till they attain a stationary value.  The number of concurrent contacts made by a node to its neighbor has been used as a parameter ($\lambda$) for describing various process of disease spread over these networks. This parametrization scheme allows us to investigate the role of network topology upon the outbreak, prevalence and final size of the epidemic.

 These simulations provide the time evolution of the dynamical state of the system and its simultaneous dependence of upon the transmission rate ($\beta$) along with the node degree($K$). From the parametric study, we note that the outbreak of the epidemic in these networks is critical in terms the number of contacts a node makes with the fraction of its neighboring nodes and there is no outbreak of epidemic for contact processes($\lambda=1$). The level of infection is quantitatively assessed from the epidemic size defined in terms of the fraction of infected individuals of the total population and their respective level of sickness. The epidemic size is determined from the final state of network. The size of the epidemic is found to increase with the parameter $\lambda$ i.e, from a contact to a reactive process. The relative epidemic spread in the various sub populations is explicitly studied to examine the microscopic criticality about the parameters in the outbreak and prevalence of disease. The understanding of this explicit dependence of epidemic size upon these parameters($\beta,K$) in the collective dynamics of the sub populations can be used for design of efficient vaccination mechanism. For a reactive process ($\lambda=\infty$), described over network with fixed number of nodes $N$ , the epidemic size is found to increase with the increase in the number of links(i.e. m=1,2,3) in the networks. The time of the spread for this process is found to have an inverse dependence upon the number of links(m values). Further the dependence of the epidemic size upon the transmission rate and degree of node are explicitly studied by projecting it in the respective parametric spaces defined as ($\rho-\beta$) and ($\rho-K$) spaces. The projection in the respective parametric spaces gives detailed dependence of epidemic size on each of these parameters. For a reactive process no critical threshold has been observed in terms of these parameters in the  outbreak of the epidemic. We concluded that this work provides a comprehensive understanding of the disease dynamics corresponding to the different spreading process over complex networks.

\acknowledgments
The authors V.S. and Y.Z. acknowledge support from the National Nature Science Foundation Committee (NSFC) of China under Project No. 61573119 and a Fundamental Research Project of Shenzhen under Project Nos. JCYJ20140417172417109, and JCYJ20140417172417090.


\newpage
\begin{center}
{\bf FIGURE CAPTIONS}
\end{center}
\noindent
Fig(1): Schematic layout to study the effect of heterogeneous transmission rate on scale free networks.

\noindent
Fig(2):Final state of the system corresponding to the different spreading processes characterized by variable $\lambda$ in the phase space defined in terms of transmission rate $\beta$ and node degree $K$. There is no outbreak of epidemic in the final state for contact process($\lambda=1$). The final states for the random walk are shown in sub plots corresponding to $\lambda=3,5$ of the variable. The final state of the system for reactive process is shown sub plot for $\lambda=\infty$.

\noindent
Fig(3): The comparative plot for the final epidemic size corresponding to different spreading processes characterized by the different values of variable $\lambda$.

\noindent
Fig(4): The comparative plots for epidemic size corresponding to random walker problems ($\lambda=3,5$) in the parametric spaces of transmission rate $\beta$ and node degree ($K$).

\noindent
Fig(5): The comparative plot for the final epidemic size for a reactive process ($\lambda=\infty$) corresponding to different values parameter $m$ of BA model.

\noindent
Fig(6): The comparative plots for epidemic size corresponding to reactive process ($\lambda=\infty$) in the parametric spaces of transmission rate $\beta$ and node degree ($K$) for different values of parameter $m$ in BA model.

\noindent
Fig(7): The comparative plot for the evolution of the epidemic size  and its corresponding evolution for the different values of parameter $m$ of BA model.

\newpage
\begin{figure}
\begin{center}
\includegraphics[angle=90, width=1\textwidth]{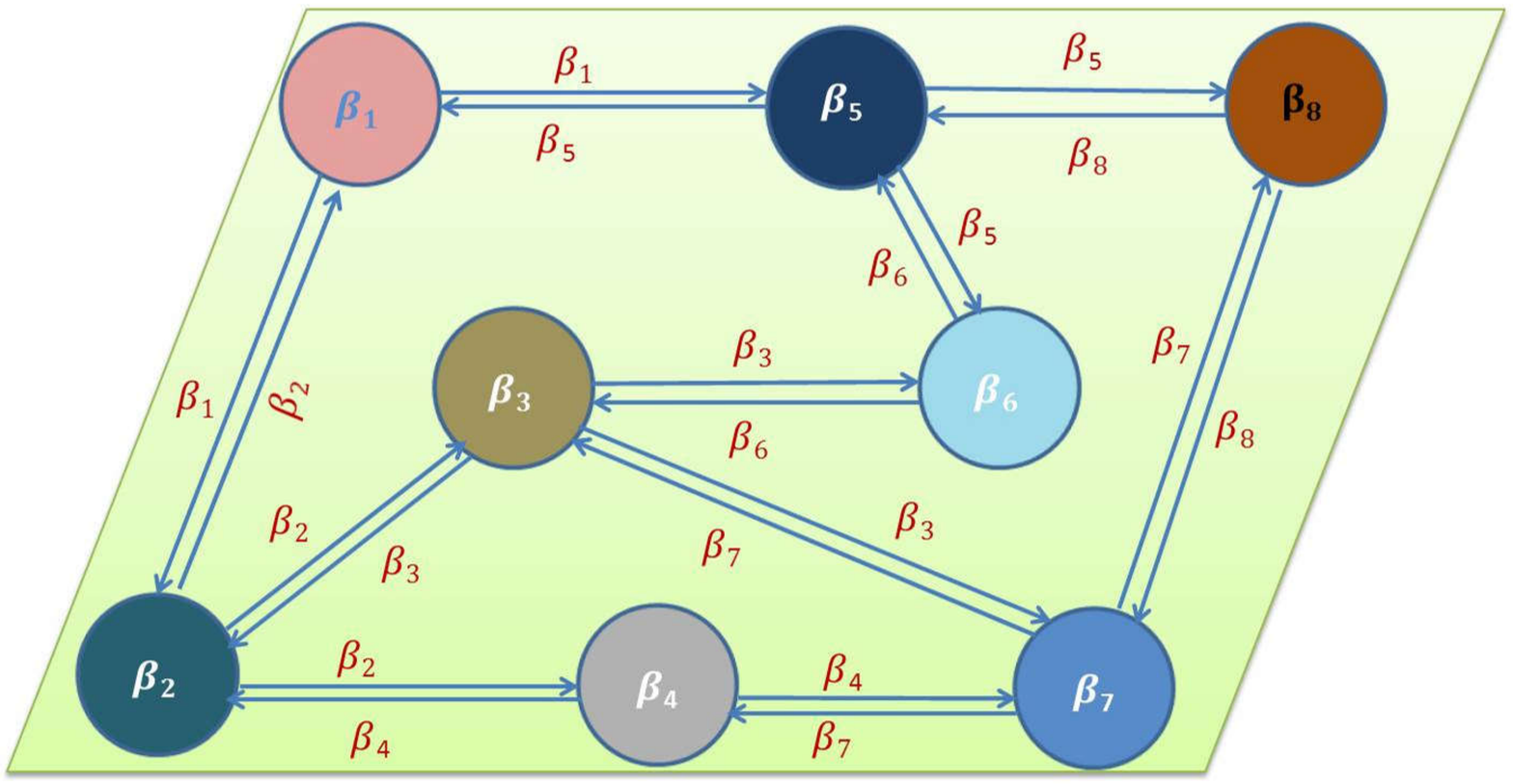}
\caption{}.
\end{center}
\end{figure}

\newpage
\begin{figure}
\begin{center}
\includegraphics[angle=0, width=1\textwidth]{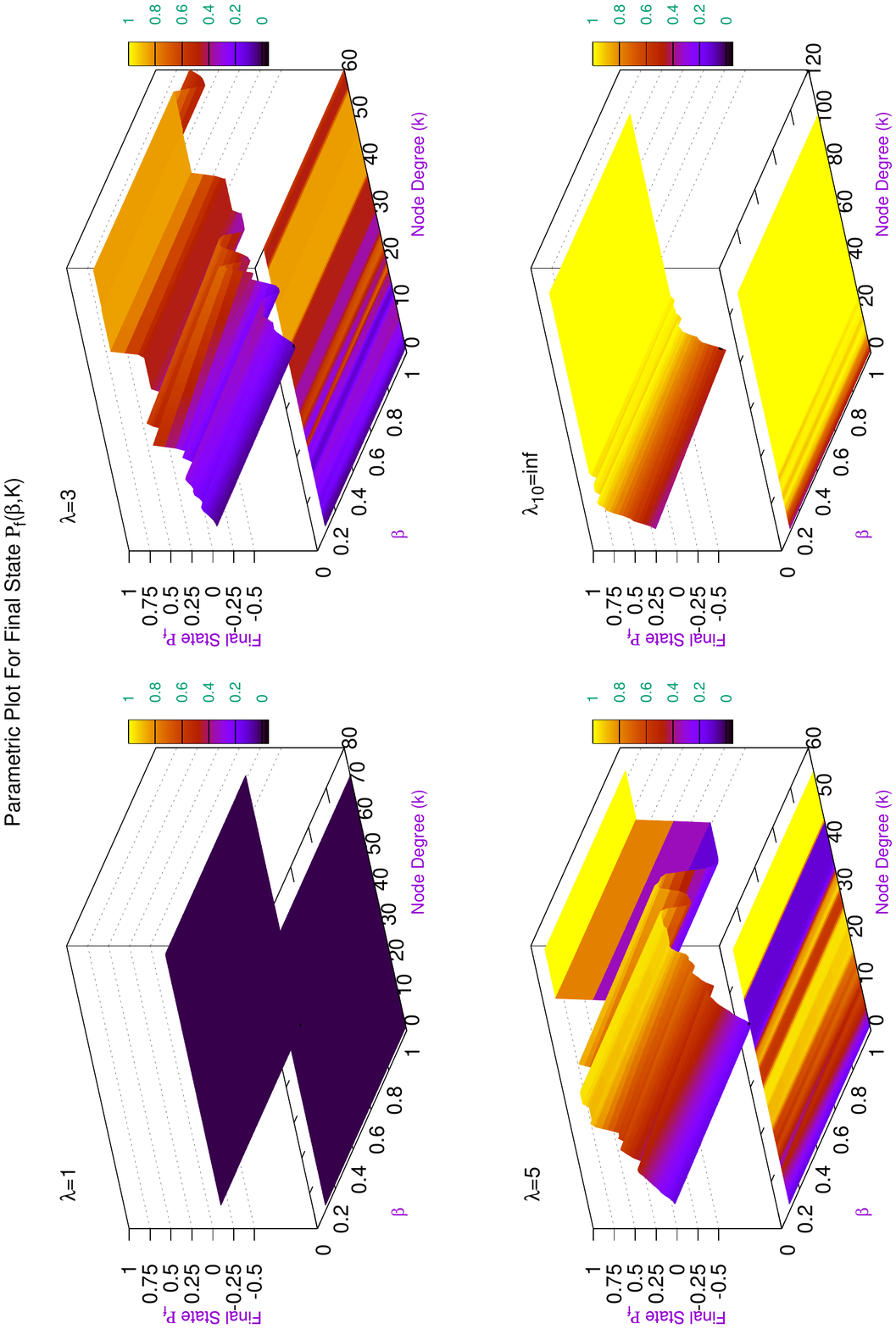}
\caption{}.
\end{center}
\end{figure}

\newpage
\begin{figure}
\begin{center}
\includegraphics[angle=0, width=1\textwidth]{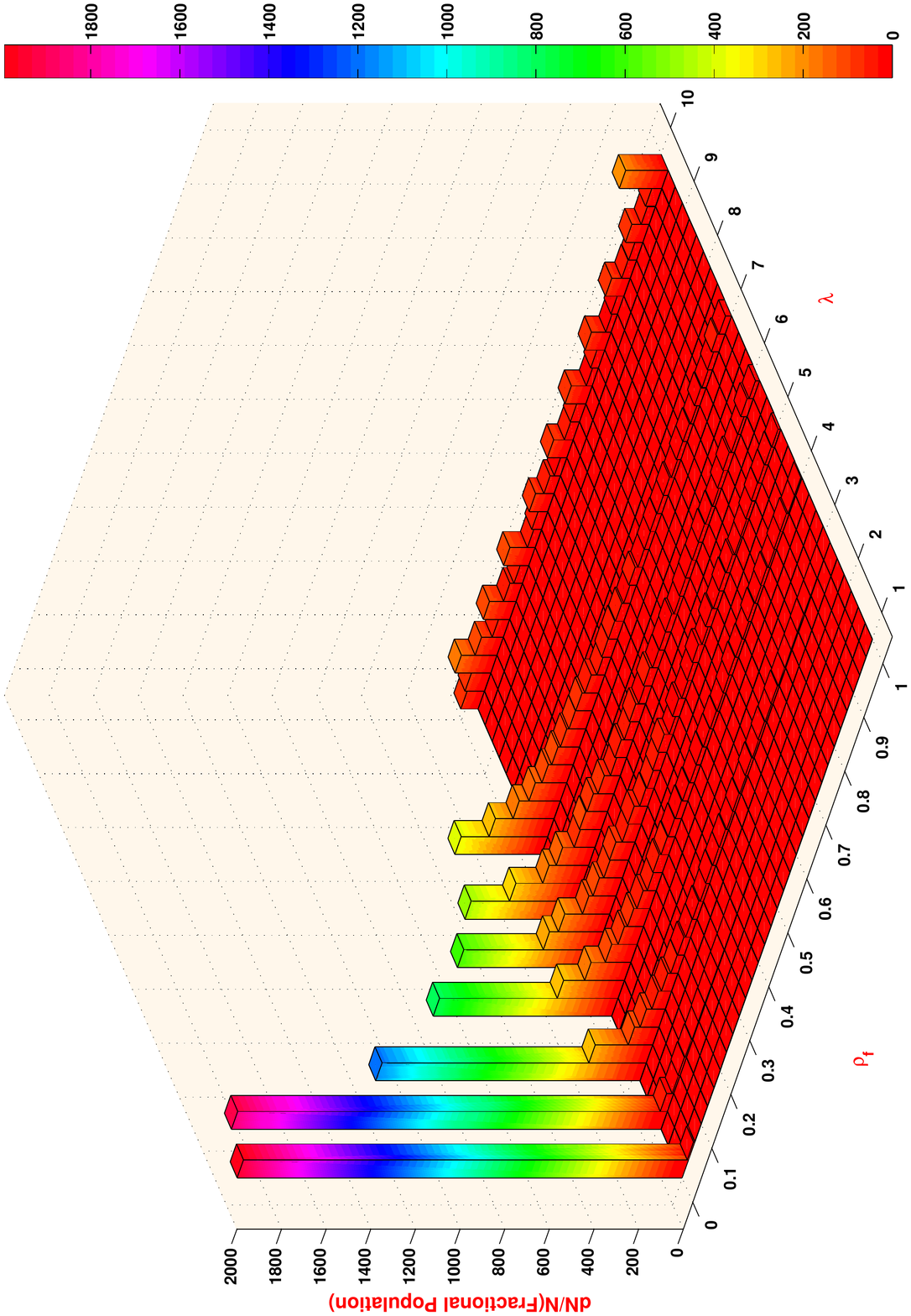}
\caption{}.
\end{center}
\end{figure}

\newpage
\begin{figure}
\begin{center}
\includegraphics[angle=0, width=1\textwidth]{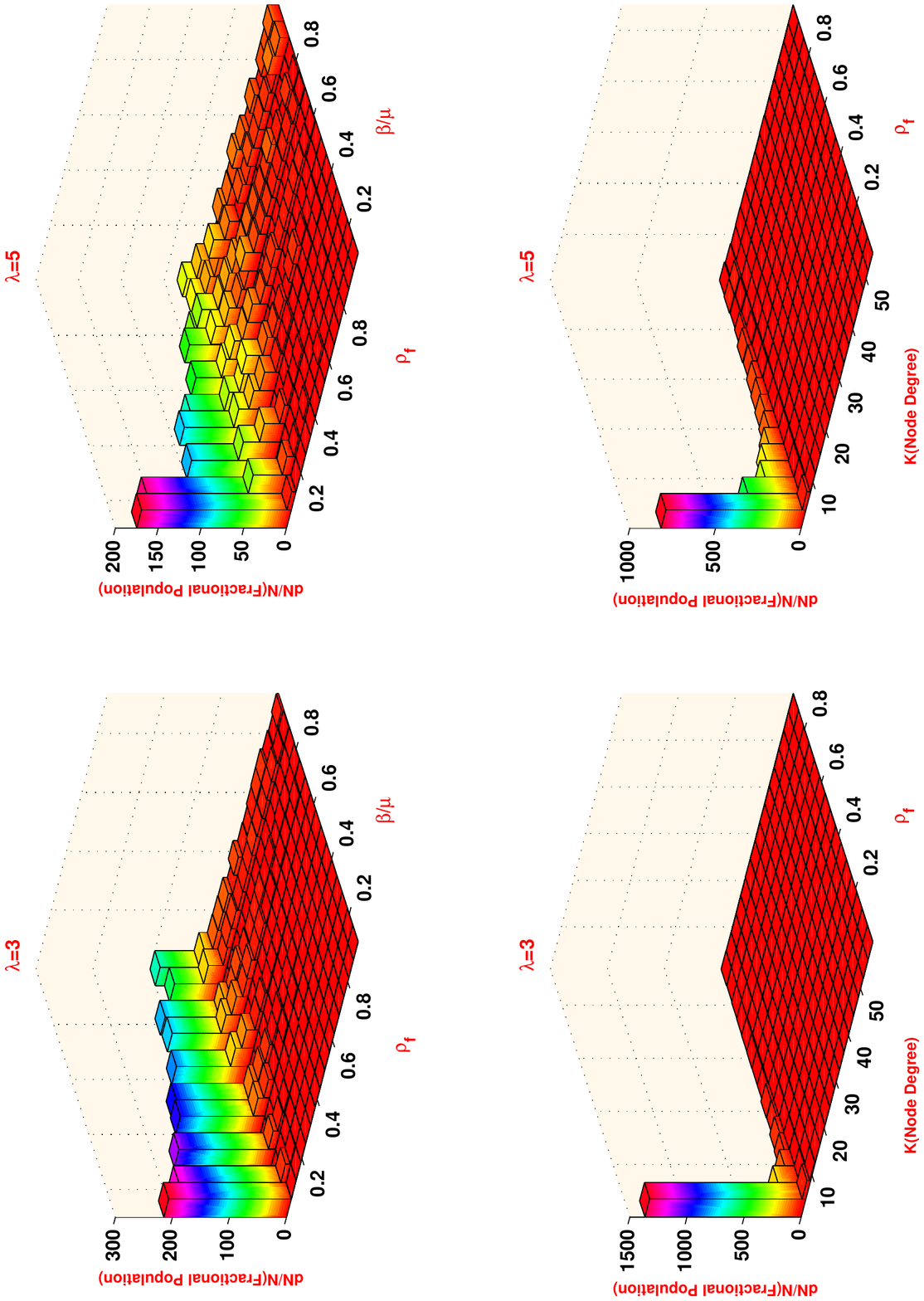}
\caption{}.
\end{center}
\end{figure}

\newpage
\begin{figure}
\begin{center}
\includegraphics[angle=0, width=1\textwidth]{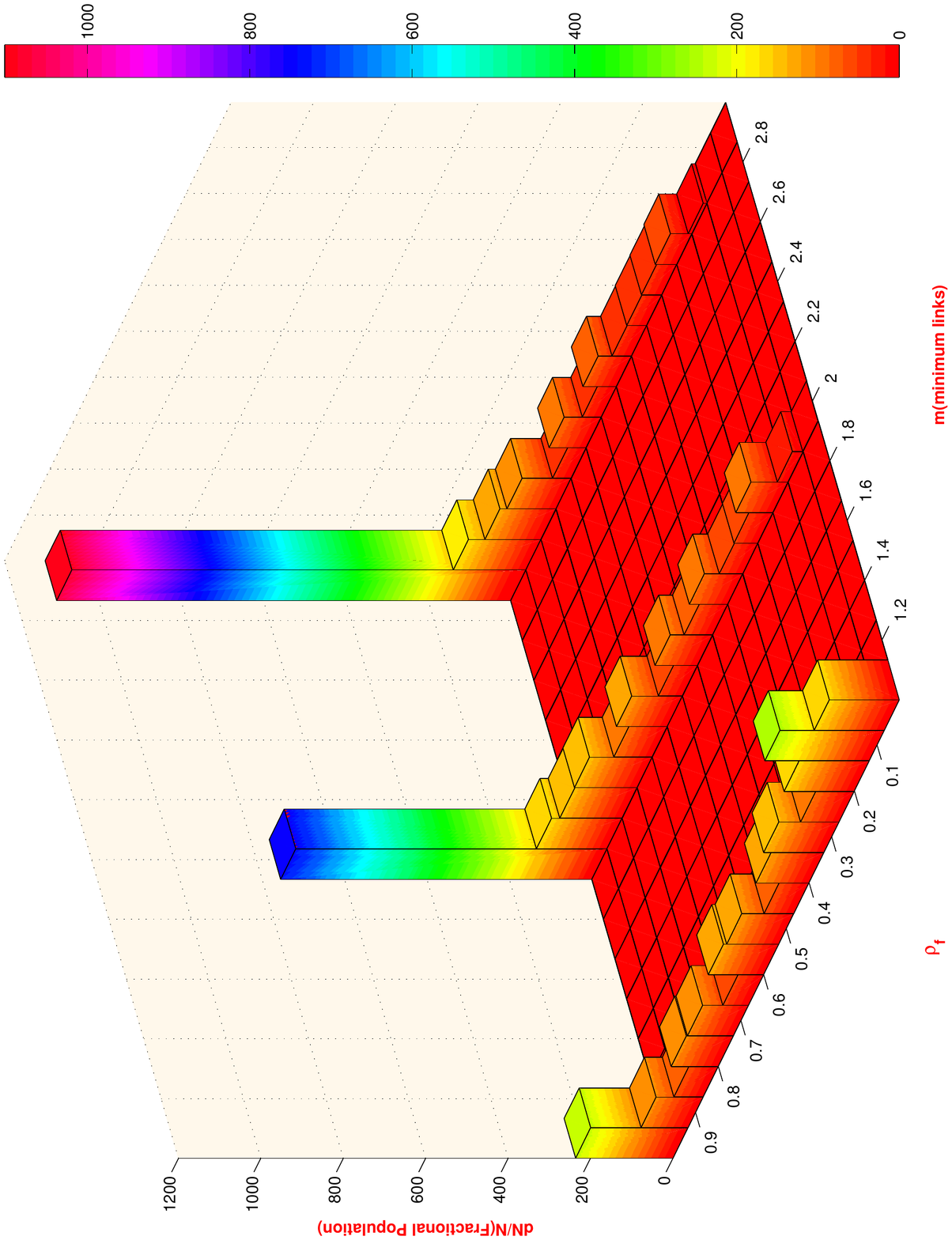}
\caption{}.
\end{center}
\end{figure}

\newpage
\begin{figure}
\begin{center}
\includegraphics[angle=0, width=1\textwidth]{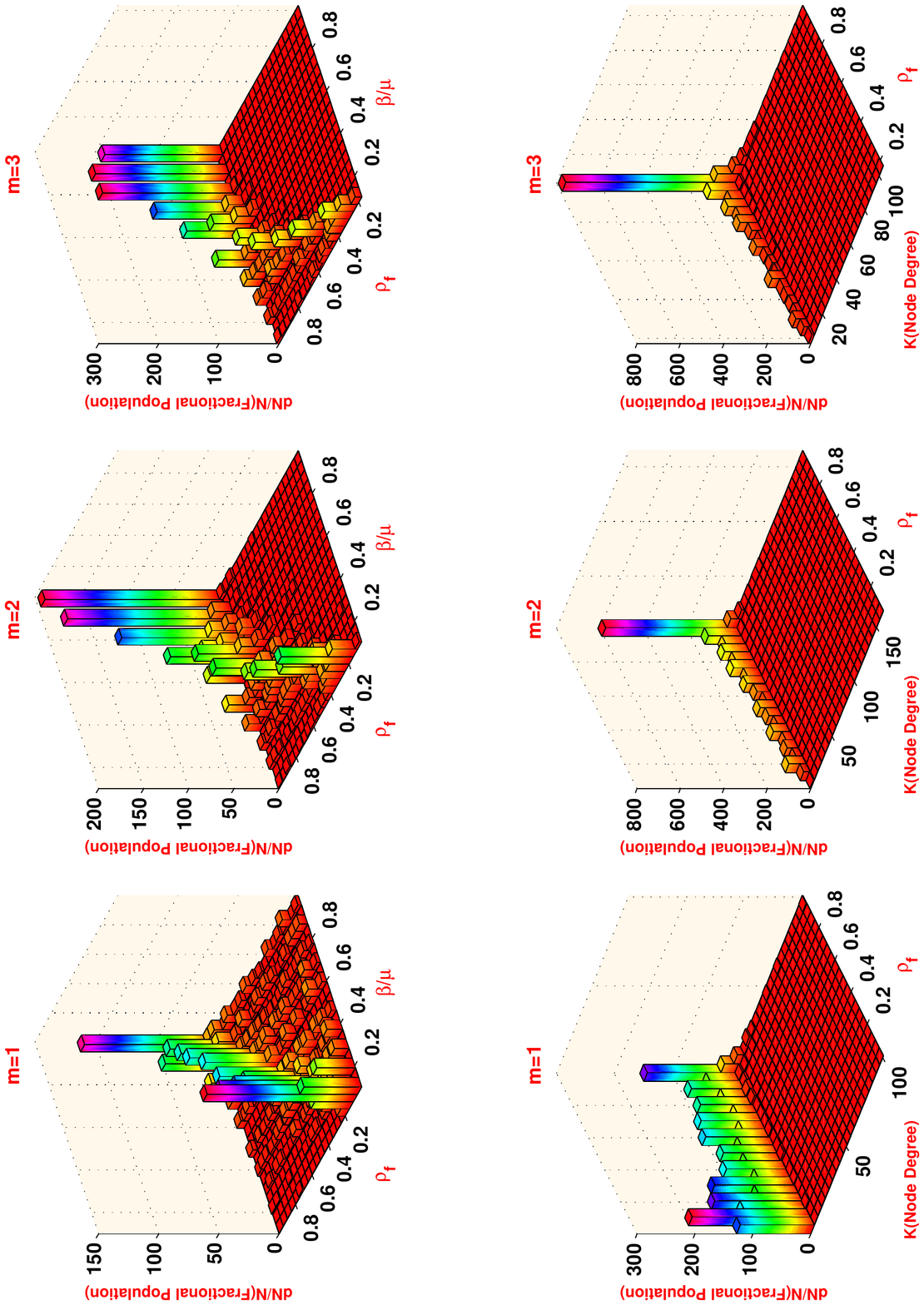}
\caption{}.
\end{center}
\end{figure}

\newpage
\begin{figure}
\begin{center}
\includegraphics[angle=0, width=1\textwidth]{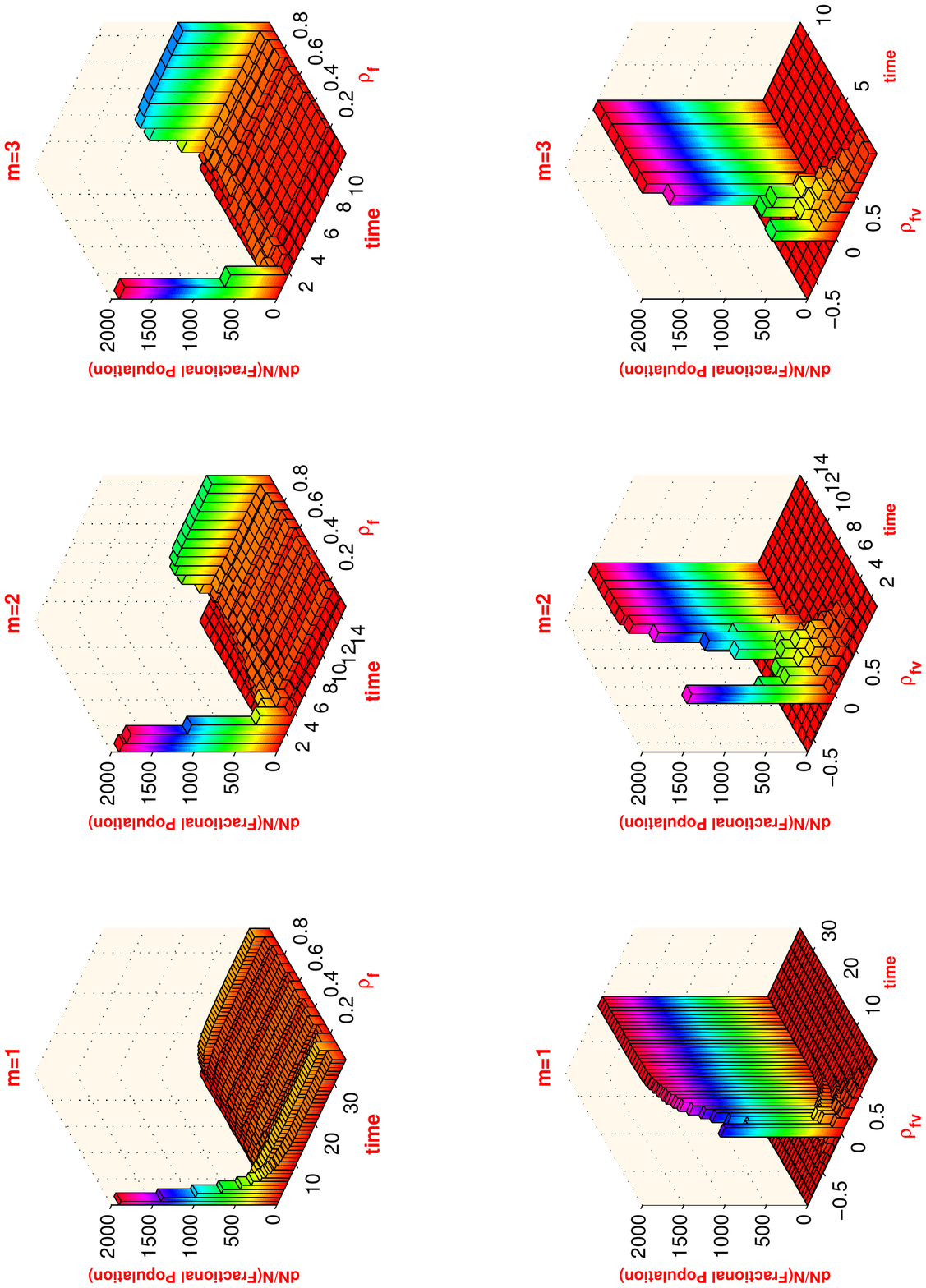}
\caption{}.
\end{center}
\end{figure}


\begin{thebibliography}{130b}
\bibitem{histro-1}
Kermack, W. O., and A. G. McKendrick, Proc. R. Soc. A 115, 700 (1927).
\bibitem{app-1}
J.O. Kephart, S.R. White, and D.M. Chess, IEEE Spectr. 30, 20 (1993).
\bibitem{app-2}
Matt J Keeling, Ken T.D Eames., R. Soc. Interface 2, 295–307 (2005).
\bibitem{app-3}
Tom Britton, Mathematical Biosciences, 225, 24–35 (2010).
\bibitem{app-4}
Stanoev A, Trpevski D, Kocarev L, PLoS ONE 9(6): e95669, (2014).
 doi:10.1371/journal.pone.0095669
\bibitem{app-5}
Romualdo Pastor-Satorras, Claudio Castellano, Piet Van Mieghem, and Alessandro Vespignani Rev. Mod. Phys. 87, 925 (2015).
\bibitem{app-6}
Cameron Nowzari, Victor M. Preciado, George J. Pappas, IEEE control systems, 36(1) (2015).

\bibitem{histro-2}
M. E. J. Newman, SIAM Rev. 45, 167 (2003).
\bibitem{histro-3}
S. Boccaletti, V. Latora, Y. Moreno, M. Chavez, and D. U. Hwang, Phys. Rep. 424, 175 (2006).
\bibitem{histro-4}
S. N. Dorogovtsev, A. V. Goltsev, and J. F. F. Mendes, Rev. Mod.Phys. 80, 1275 (2008).
\bibitem{histro-5}
 M. E. J. Newman, Networks: An Introduction. Cambridge, U.K.: Cambridge Univ. Press, (2010).


\bibitem{math-1}
N.T.J. Bailey, The Mathematical Theory of Infectious Diseases, 2nd ed. (Griffin, London, 1975).
\bibitem{math-2}
J.D. Murray, Mathematical Biology (Springer Verlag, Berlin, 1993).
\bibitem{math-3}
 H.W. Hethcote, SIAM Rev., vol.42, no. 4, pp. 599–653, (2000).



\bibitem{intro-8}
Gonzalo M. Vazquez-Prokopec, T. Alex Perkins, Lance A. Waller, Alun L. Lloyd, Robert C. Reiner Jr, Thomas W. Scott, Uriel Kitron, Trends in Parasitology, Vol. 32, Issue 5, 356–367 (2016).


\bibitem{BA-1}
A.-L. Barabasi and R. Albert, Science 286, 509 (1999).
\bibitem{BA-2}
A.-L.Barabasi, R. Albert, and H. Jeong, Physica A 272, 173 (1999).
\bibitem{SF-1}
G. Caldarelli, Scale-Free Networks (Oxford University Press,Oxford, 2007).
\bibitem{SF-2}
R. Pastor-Satorras and A. Vespignani, Evolution and Structure of the Internet: A Statistical Physics Approach (Cambridge
University Press, Cambridge, UK, 2004).
\bibitem{SF-3}
F. Liljeros, C. R. Edling, L. A. N. Amaral, H. E. Stanley, and Y. Aberg, Nature (London) 411, 907 (2001).
\bibitem{SIS-1}
Pastor-Satorras,R. and Vespignani,A., Phys. Rev. Lett.,86, 3200 (2001).
\bibitem{SIS-2}
Pastor-Satorras,R.and Vespignani,A.,Phys. Rev. E, 63, 066117 (2001).
\bibitem{SIS-3}
M. Barthelemy, A. Barrat, R. Pastor-Satorras, and A. Vespignani,Phys. Rev. Lett. 92, 178701 (2004).
\bibitem{SIS-4}
J. Gomez-Gardenes, V. Latora, Y. Moreno, and E. Profumo,Proc. Nat. Acad. Sci. USA 105, 1399 (2008).
\bibitem{SIS-5}
Newman M. E. J., Phys. Rev. E, 66, 016128 (2002).
\bibitem{SIS-6}
S. Gomez, A. Arenas, J. Borge-Holthoefer, S. Meloni, and Y. Moreno, Europhys. Lett. 89, 38009 (2010).
\bibitem{SIS-7}
Sergio Gomez, Jesus Gomez-Gardenes, Yamir Moreno, and Alex Arenas, Phys. Rev. E 84, 036105 (2011).
\bibitem{SIS-8}
Daniel Smilkov and Ljupco Kocarev, Phy. Rev. E, 85, 016114 (2012).

\end{thebibliography}
\end{document}